\documentclass[letter,twocolumn,showpacs,aps]{revtex4}

\expandafter\let\csname equation*\endcsname\relax
\expandafter\let\csname endequation*\endcsname\relax

\usepackage{amsmath}
\usepackage{amssymb}
\usepackage{color}
\usepackage{verbatim}
\usepackage{array}
\usepackage{graphicx}
\usepackage{epsfig}
\usepackage{bm}
\usepackage[ragged]{sidecap}
\usepackage{dsfont}
\usepackage{ulem}
\usepackage{mathrsfs}
\usepackage{setspace}

\usepackage{subfigure}

\newcommand{\ds}{\displaystyle}

\newcommand{\dd}{\mathrm{d}}

\newcommand{\inn}[1]{\langle #1 \rangle}

\begin{document}
\preprint{\hfill\parbox[b]{0.3\hsize}{ }}

\title{A Fabry-Perot interferometer with quantum mirrors: \\ nonlinear light transport and rectification}

\author{F. Fratini$^{1,2}$\footnote{
\begin{tabular}{ll}
E-mail addresses:& ffratini@fisica.ufmg.br\\
&fratini.filippo@gmail.com
\end{tabular}},
E. Mascarenhas$^1$, L. Safari$^{3,4}$, J-Ph. Poizat$^2$, D. Valente$^5$, A. Auff\`{e}ves$^2$, D. Gerace$^6$, and M. F. Santos$^1$}
\affiliation
{\it
$^1$Departamento de F\'isica, Universidade Federal de Minas Gerais, CP 702, 30123-970 Belo Horizonte, Brazil \\
$^2$Institut N\'eel-CNRS, 25 rue des Martyrs, BP 166,  38042 Grenoble Cedex 9, France\\
$^3$Department of Physics, University of Oulu, Box 3000, FI-90014 Oulu, Finland\\
$^4$IST Austria, Am Campus 1, A-3400 Klosterneuburg, Austria\\
$^5$Instituto de F\'isica, Universidade Federal de Mato Grosso, Cuiab\'a MT, Brazil\\
$^6$Dipartimento di Fisica, Universit\`a di Pavia, via Bassi 6, I-27100 Pavia, Italy
}

\date{\today}

\begin{abstract}
Optical transport represents a natural route towards fast communications, and it is currently used in large scale data transfer. The progressive miniaturization of devices for information processing calls for the microscopic tailoring of light transport and confinement at length scales appropriate for the upcoming technologies. With this goal in mind, we present a theoretical analysis of a one-dimensional Fabry-Perot interferometer built with two highly saturable nonlinear mirrors: a pair of two-level systems. 
Our approach captures non-linear and non-reciprocal effects of light transport that were not reported previously. Remarkably, we show that such an elementary device can operate as a microscopic integrated optical rectifier. 
\end{abstract}

\pacs{42.50.Nn, 42.50.St, 42.60.Da, 42.65.-k}

\maketitle

{\it Introduction: }
There is a growing interest in the realization of quantum optical systems in which single emitters are strongly coupled to one-dimensional (1D) radiation modes for efficient light transport \cite{Shen2005,Chang,Shen2007} and quantum information processing \cite{Daniel2012}. The ultimate goal would be to progressively replace or hybridize current microelectronics with integrated optical devices in order to enhance data capacity, transmission velocity, and efficiency. One of the benchmarks for information processing is the ability to control the directionality of energy flux within a specific system architecture, a task that generically requires 1D propagation channels and nonlinear components. 
Furthermore, so-called rectifying devices provide the unidirectional isolation of strategical centres in electronic circuits.
The combination of these properties has allowed for the technological revolution of microelectronic processors in the last century, and a similar development for the transport of light is necessary if one is to expect photon-based computing systems. 
In this letter, we show how two-level quantum systems may be employed as non-linear mirrors forming a Fabry-Perot interferometer. Optical rectification is a direct consequence of the nonlinear nature of such interferometer. If integrated within an optical circuit, this rectifying device would prevent unwanted signals (or noise) to travel back, thus preserving the processing capabilities at the source. This is of utmost importance in the quantum regime, e.g. to prevent decoherence at the sender of the signals.

Several experiments have already demonstrated the combination of strong nonlinear behavior and 1D light propagation in different system implementations, such as trapped ions coupled to focused light beams or optical fibers \cite{Het2011,Mon2013}, superconducting circuits coupled to microwave transmission lines \cite{Abdu2010,VanL2013,Lal2013}, and semiconductor quantum dots or vacancy (e.g., N-V) centers coupled to photonic or plasmonic waveguides \cite{lodhal2008,strain2014,lukin2007,huck2011}. Among these attempts, the use of solid state quantum emitters as artificial two-level systems (TLS) is specially promising due to their nanoscale dimensions, their extreme nonlinear properties, and their tunability, thanks to the use of external electrostatic gates \cite{Gate}, applied magnetic fields \cite{MagnQD, MagnQD2}, or mechanical strain \cite{strain2014}. These combined advantages have led to a wide range of theoretical proposals and recent experiments, with the aim of building single photon emitters \cite{Gr1}, single-photon light switches and transistors \cite{Chang}, quantum optical diodes \cite{Roy2010,Shen2011,Dudu} and interferometers \cite{Dar2009}. Following these proposals, a pair of TLS coupled to a 1D waveguide can be expected as one of the simplest configurations where tunable non-linear and non-reciprocal optical phenomena at the quantum level could be practically realized, besides allowing for photonic mediated interactions between distant qubits \cite{Kie2005,Bus2013,Bar2013}. 

In this work we employ a semi-classical analysis to theoretically treat the transport of light in a \textit{Quantum Fabry-Perot} (QFP) interferometer built from two TLS embedded in a 1D photonic channel, drawing inspiration from the recently demonstrated analogy between a single TLS and an optical mirror~\cite{Het2011}. After validating our theoretical approach through comparison with previous results based on a similar model for the case of two identical TLS \cite{Shen2005,Lal2013,Bar2013}, we thoroughly study the case of two different TLS. In this latter case, we show that the QFP interferometer manifests non-reciprocal effects, not captured in previous works, that enable to rectify light transport through the 1D channel. Remarkably, we find regions where both light rectification and transmission exceed 92\%, depending on the system parameters. In our approach, the TLS are treated as quantum systems, and rectification emerges out of their highly nonlinear behavior, while the light field is treated as a classical input. Given the generality of this method in describing light transport within this QFP interferometer, our results can be adapted to a number of different physical implementations, as discussed at the end of the letter. Differently from previous proposals 
\cite{Shen2011,Yu2008,Tian2012,Lira2012}, our QFP interferometer does not require the application of external fields to produce non-linear effects on light transport.

\begin{figure}[t]
\centering
\includegraphics[scale=0.5,clip=true,trim=0cm 0 0cm 0]{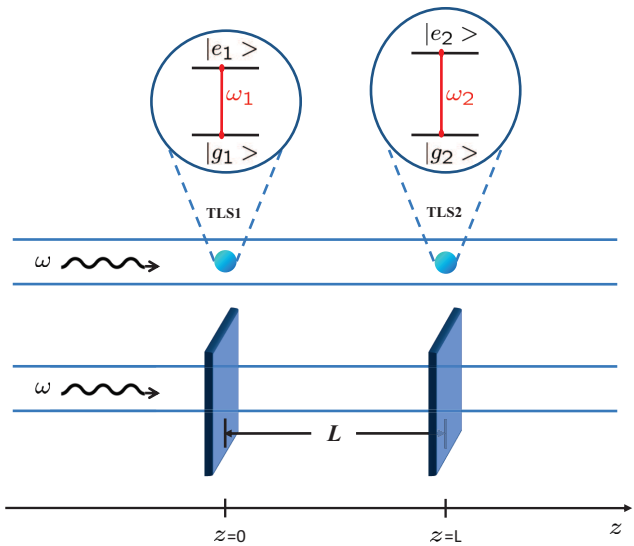}
\caption{(Color online). A pair of two-level quantum systems in a one-dimensional waveguide as a quantum Fabry-Perot interferometer, and its classical counterpart.}
\label{fig:Fig1}
\end{figure}

\medskip

{\it Fabry-Perot model: }
We consider two TLS embedded in a 1D waveguide, as shown in Fig. \ref{fig:Fig1}. Light with angular frequency $\omega$ and power $p_{inc}$ is injected into the waveguide. We shall use the terms ``intensity'' and ``power'' interchangeably throughout the manuscript. For these, we will use the symbol $p$, which is a dimensionless quantity representing the number of photons per lifetime. We denote with TLS1 (TLS2) the first (second) quantum emitter lying on the light path, if light is shined from left-to-right as in Fig. \ref{fig:Fig1}. The TLS1 (TLS2) has transition frequency $\omega_{1(2)}$, decay rate $\gamma_{1(2)}$, position $z=0$ ($z=L$). The detuning of the incoming light with respect to the TLS1 (TLS2) transition frequency is $\delta\omega_1=\omega-\omega_1$ ($\delta\omega_2=\omega-\omega_2$).
Within a semi-classical approach, we treat such a system in analogy to a Fabry-Perot interferometer (see Fig. \ref{fig:Fig1}), where the reflectances of the mirrors are given by the reflectances of the TLS, as obtained in a quantum mechanical framework. 
These latter can be readily derived from Ref. \cite{Daniel} (see \cite{Suppl}):
\begin{equation}
\label{eq:RQDs}
R_{1(2)}=\frac{\gamma_{1(2)}^2}{\gamma_{1(2)}^2+4\delta\omega_{1(2)}^2+4p_{1(2)}\gamma_{1(2)}^2}~,
\end{equation}
where $p_{1(2)}$ is the power impinging onto the TLS1 (TLS2), i.e. $\gamma_{1(2)}p_{1(2)}$ is the number of photons per second impinging onto TLS1 (TLS2). 
The quantity $R_{1(2)}$ represents the fraction of light power that TLS1 (TLS2) reflects back into the 1D channel.
Furthermore, $\theta_{1(2)}=\arctan\left[{2\delta\omega_{1(2)}}/{\gamma_{1(2)}}\right]-\pi$ is the phase-shift given by the TLS1 (TLS2) to the light upon each reflection \cite{Bar2013}. The phase-shift given by either TLS to the transmitted light is neglected, as usual for mirrors. 

The fraction of light power that the FP interferometer transmits, i.e. the FP transmittance, can be calculated as \cite{Suppl}
\begin{equation}
\label{eq:T}
T=\ds \frac{1}{F_1+F_2\sin^2(2\mu+\theta_+)}~,
\end{equation}
where
\begin{equation}
F_1=\frac{(1-\sqrt{R_1R_2})^2}{(1-R_1)(1-R_2)}\;,\;
F_2=\frac{4\sqrt{R_1R_2}}{(1-R_1)(1-R_2)}\;,
\end{equation}
while $\mu=n\omega L/(2c)$, $n$ is the effective refractive index of the waveguide, $c$ is the speed of light in vacuum and $\theta_+=(\theta_1+\theta_2)/2$.
In order to use Eq. \eqref{eq:T}, we need first to find what the values for $R_1$ and $R_2$ are. This reduces to the question of finding what the values for $p_1$ and $p_2$ are. These latter can be obtained by numerically solving a system of coupled equations. The details of such a calculation are given in the supplemental material \cite{Suppl}. 
Thus, by using Eqs. \eqref{eq:T} and \eqref{eq:RQDs}, together with the numerical values for $p_{1,2}$, the transmittance $T$ can be numerically calculated for any set of the (externally adjustable) variables $\gamma_{1,2}$, $\delta\omega_{1,2}$, $L$, $p_{inc}$.

\medskip

{\it Non-linear light transport: }
The semiclassical approach employed in this work has been fully validated by comparing our results to solutions based on quantum mechanical models taken from the literature \cite{Bar2013, Shen2005, Lal2013} (see \cite{Suppl}). In particular, we stress that the present approach allows to calculate the light transport for any incident light power, as well as for different atomic frequencies and decay rates.

\begin{figure}[t]
\centering
\includegraphics[scale=0.48]{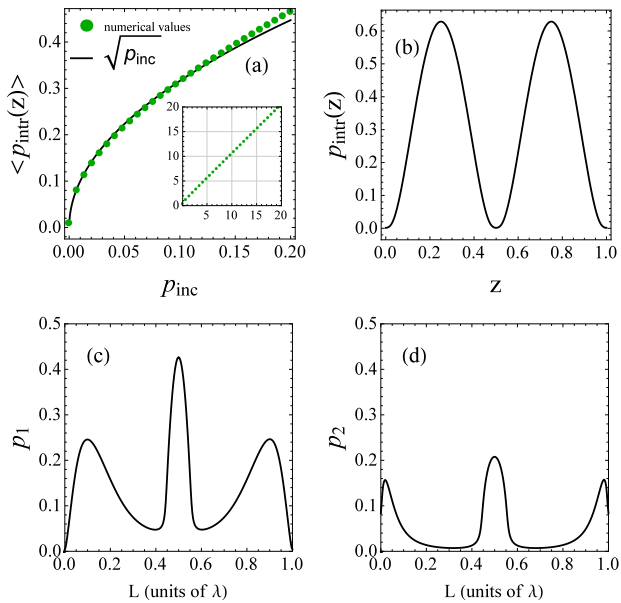}
\caption{(color online). 
(a) Average intracavity intensity. The inset shows the same quantity for larger values of the abscissa; (b) Intracavity intensity;
(c) Intensity impinging onto TLS1; (d) Intensity impinging onto TLS2. 
$p_{inc}$ is the incident power, while $L$ is the TLS distance in units of photon wavelength. 
$\delta\omega_1=\delta\omega_2=0$ in all panels. 
$L=1$ in (a) and (b). $p_{inc}=0.1$ in (b), (c) and (d). 
}
\label{fig:Fig2}
\end{figure}

First, we explore the light intensity between the TLS (intracavity intensity) and at the TLS positions, respectively. For simplicity, in the following we will consider  $\gamma_1=\gamma_2=1\equiv\gamma$ and $L$ in units of the photon wavelength, $\lambda=2\pi c/(n\omega)$. In a standard FP interferometer, large intracavity intensity is present when the mirror reflectances are close to 1. 
In line with our analogy, high intracavity intensity is expected in the present model when the TLS reflectances $R_{1,2}$ are nearly 1, which is the case when light is shined in resonance with the TLS and at low incident power. 
Let us denote by $p_{intr}(z)$ the intracavity intensity at the position $z$, where $0<z<L$, and by $\langle p_{intr}(z)\rangle$ the average intracavity intensity: $\langle p_{intr}(z)\rangle=\int_0^L p_{intr}(z) \dd z/L$.
In Figs. \ref{fig:Fig2}(a) and (b), these two quantities are plotted as a function of $p_{inc}$ and $z$, respectively. From panel (a) we notice that the relation between $\langle p_{intr}(z)\rangle$ and $p_{inc}$ is {\it non-linear}. In fact, by supposing low incident power and $\delta\omega_1=\delta\omega_2=0$, it can be analytically shown that the average intracavity intensity is well approximated by $\langle p_{intr}(z)\rangle\approx\sqrt{p_{inc}}$. In Fig. \ref{fig:Fig2}(a), such approximate expression and the exact numerical values for $\langle p_{intr}(z)\rangle$ are directly compared. The relation $\langle p_{intr}(z)\rangle\approx\sqrt{p_{inc}}$ indeed yields $\langle p_{intr}(z)\rangle\gg p_{inc}$, as we expected from the discussion above. Furthermore, this non-linear relation marks a stark difference with respect to the standard Fabry-Perot interferometer, where a linear relation between incident and intracavity intensities holds \cite{Yariv}. 

In our model, only for large $p_{inc}$ the average intracavity intensity can be well approximated by a linear function of $p_{inc}$ (see panel (a), inset). Specifically, for large $p_{inc}$ the average intracavity intensity asymptotically satisfies the relation $\langle p_{intr}(z)\rangle\approx p_{inc}$, as expected.

Finally, for low incident power ($p_{inc}\lesssim 1$), light between the atoms forms a standing wave where nodes are present (see panel (b)). Nodes are at positions $z=0$, $1/2$ and $1$, as one would expect for a cavity with perfect mirrors.

It is instructive to show the light intensities at the sites of the TLS as the distance $L$ varies, while $p_{inc}$ is kept constant. In Figs. \ref{fig:Fig2}(c) and (d), we plot $p_1$ and $p_2$  for incident power $p_{inc}=0.1$ and $\delta\omega_1=\delta\omega_2=0$. We notice that at $L=0$, $1$ the light intensity at the TLS1 position is identically 0 and, consequently, the light intensity at the TLS2 position equals the incident intensity. This remains true up to about $p_{inc}\lesssim 1$, and it is caused by the fact that the back reflected light from the TLS2 turns out to be $\pi$ shifted with respect to the incident light, at the site of the TLS1 (see \cite{Suppl} for more details). On the other hand, at $L=0.5$ the light intensity at the TLSs positions is maximal.
\medskip

{\it Rectification: }
The joint implementation of TLS and 1D waveguides is believed to represent the future building-blocks of nanoscale optoelectronics \cite{Duan2001,Seth2009}. The realization of nanoscale devices that allow unidirectional light transmission is of utmost importance in this field, and is thus subject of current research \cite{Bi2011,Fan2012}. However, most of the attempts to realize or propose optical diodes able to work at the quantum regime lack real miniaturization possibilities and control at the nanoscale \cite{Shen2011,Yu2008,Tian2012,Lira2012,Dudu,Cheng2012,Wang2013,Roy2010}.
Here we show that two TLS embedded in a 1D waveguide provide the requested features for building a microscopic and integrable optical diode. The realization of this quantum optical diode is feasible with the state-of-the-art technology, as discussed in the following section.

\begin{figure}[t]
\centering
\includegraphics[scale=0.35]{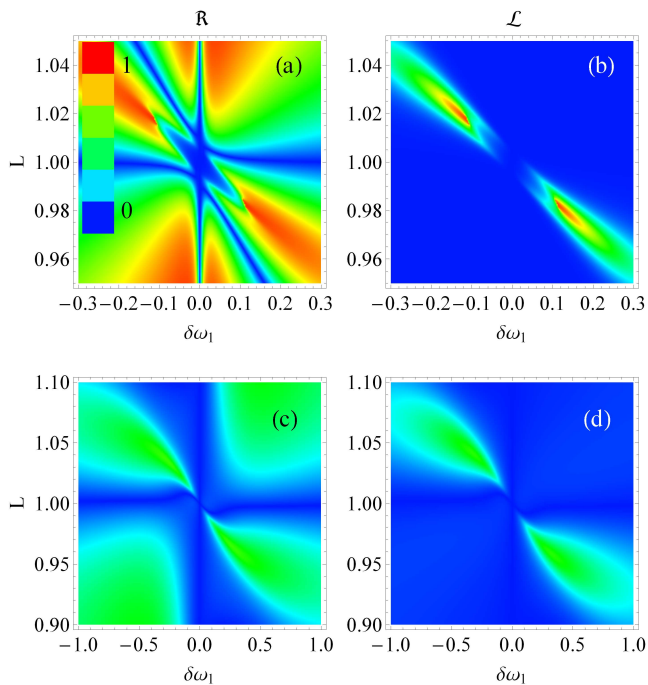}
\caption{(color online). 
Light rectification. Parameters $\mathcal{R}$ and $\mathcal{L}$ are plotted. $\delta\omega_2=0$ in all panels. $p_{inc}=0.001$ in (a) and (b). $p_{inc}=0.1$ in (c) and (d). 
}
\label{fig:Fig3}
\end{figure}

We define the rectifying factor for an optical diode as \cite{Dudu,Casati}
\begin{equation}
\mathcal{R}=\ds\frac{\big| T_{12}-T_{21} \big|}{T_{12}+T_{21}}~,
\end{equation}
where $T_{12}$ is the transmittance for the case light is shined from left-to-right (as in Fig. \ref{fig:Fig1}), while $T_{21}$ is the transmittance in the optical inverse situation where light is shined from right-to-left. We shall take $\mathcal{R}$ and 
$\mathcal{L}=T_{12}\mathcal{R}$ as figures of merit to quantify the non-reciprocal effects that our microscopic FP manifests. In Fig. \ref{fig:Fig3}, the quantities $\mathcal{R}$ and $\mathcal{L}$ are shown as functions of $L$ and $\delta\omega_1$, while $\delta\omega_2\approx0$. In panels (a) and (b), we investigate the case $p_{inc}=0.001$, which may be considered equivalent to the single-photon regime (see Fig. S2 of \cite{Suppl}). High levels of light rectification and transmission are evident. Specifically, some areas in the color scale plot are characterized by both $\mathcal{R}$ and $\mathcal{L}$ greater than 0.92.  By increasing the incident power, these areas broaden, while $\mathcal{R}$ and $\mathcal{L}$ decrease (see panels (c) and (d) where the same quantities are plotted for $p_{inc}=0.1$). In (c) and (d), the highest values for $\mathcal{R}$ and $\mathcal{L}$ are $\approx0.53$ and $0.52$, respectively. 

High values for $\mathcal{R}$ and $\mathcal{L}$ in Figs. \ref{fig:Fig3}(a) and (b), could be understood as follows. Light is in resonance with TLS2 and at low power, while it is in general not in resonance with TLS1, unless we are in the central region of the plots where $\delta\omega_1=0$. Under such conditions, we have $R_1<1$ and $R_2\approx1$. When light is incident from right-to-left, it encounters TLS2 first, which implies full reflection (being $R_2\approx1$). 
On the other hand, when light is incident from left-to-right, it encounters TLS1 first,  hence a significant amount of that light is coherently transmitted to TLS2 (since $R_1<1$). Then, TLS2 totally reflects such radiation back into the 1D channel to TLS1. Such light acquires a phase-shift that depends on $L$, due to the path length. 
At this point, TLS1 must deal with both the phase-shifted light coming back from TLS2 and the incident light that is forwardly directed. Both are partially reflected and partially transmitted. However, since light is not in resonance with TLS1, the light reflected from TLS1 acquires a further phase-shift $\theta_1$ that depends on $\delta\omega_1$ (see after Eq. \eqref{eq:RQDs}).  The two phase-shifts, the one depending on $L$ and the one depending on $\delta\omega_1$, can give constructive or destructive interference. 
For some values of $L$ and $\delta\omega_1$, we get destructive interference for light exiting the FP from the left, while constructive interference for light directed toward TLS2. 
Those values provide high level of rectification showed in Fig. \ref{fig:Fig3}(a). In Figs. \ref{fig:Fig3}(c) and (d), both $R_1$ and $R_2$ are considerably lower than $1$, since here the incident light power is not very low. By re-applying the foregoing discussion, we expect and find lower degree of light rectification. 

We finally point out that the results shown in Fig. \ref{fig:Fig3} do not change significantly within the interval $-0.01\lesssim\delta\omega_2\lesssim 0.01$. For configurations where none of the two TLS is in resonance with the incident light beam, there is no region where both $\mathcal{R}$ and $\mathcal{L}$ simultaneously display large values.

\medskip

{\it Physical implementation: }
The QFP interferometer introduced in this work can be implemented in a number of different technologies and material platforms. 
In particular, we outline three main architectures as promising candidates to observe such non-reciprocal behavior. 

First, superconducting circuits have emerged in the last few years as an outstanding platform to realize quantum optical functionalities in the microwave range. In this respect, the QFP interferometer can be realized with the state-of-art technology. Considering recent experiments \cite{VanL2013}, we notice that the 
system parameters for attaining maximal light rectification and transmission are well within reach. Although it does not represent a miniaturized version of our proposed device, such microwave circuit implementation of the QFP interferometer is likely to be the most promising candidate for a first proof-of-principle demonstration of the rectifying features, also thanks to the high level of electrostatic control on state of the single superconducting qubits as TLS.

As a second alternative, we notice that remarkable progress has been lately achieved in coupling semiconductor quantum dots to 1D photonic wires \cite{lodhal2008,Gr1} or to semiconducting micro-pillars \cite{lanco2012}. Such artificial atoms behave as almost ideal TLS, and growing stacks of two or more QDs along the same axis and at distances on the order of the optical emission wavelength ($\sim 1$ $\mu$m) is at the level of current technology \cite{badolato2005}. Moreover, such a nanophotonic platform would naturally represent an fully integrated quantum optical version of our proposed device.

Finally, NV (Nitrogen vacancy) centers in diamond coupled to 1D surface plasmons \cite{huck2011} can be an interesting possibility to implement a QFP model. In this case, large values of the light-matter coupling rate could be achieved, owing to the strong confinement of plasmonic modes close to the metallic nanowire surface. This could allow to easily achieve the requested parameters range for light rectification along the 1D axis, i.e. large phase shifts produced by each TLS on incoming light. 

The material and engineering efficiency in preparing the 1D system is standardly quantified by an efficiency parameter $\beta$ ranging from $0$ (minimal efficiency) to $1$ (maximal efficiency). $\beta$ quantifies the strength of the TLS-light coupling in the 1D material. 
Remarkably high values of $\beta$ have been attained in recent experiments: in superconducting circuits $\beta\approx0.99$ \cite{VanL2013}, while in semiconductor quantum dots coupled to photonic wires \cite{Mun2013} or to photonic crystals \cite{Han2008} $\beta\approx 0.95, 0.89$, respectively.




\medskip

{\it Summary and conclusions: }
We modeled a pair of two-level quantum systems embedded in a one-dimensional waveguide as a Fabry-Perot quantum interferometer, where the two quantum systems play the role of highly saturable and nonlinear mirrors. Beside manifesting non-linear effects, we showed that this quantum interferometer can work as a very efficient integrated optical diode, with unprecedented figures of merit in terms of simultaneous light rectification and transmission, and thus with potential applications in integrated optical photonics.
Such a quantum optical diode can be implemented with several integrated one-dimensional designs employing different state-of-the-art technologies and materials, and dimensions ranging from nanometer to millimeter sizes. Unconditional quantum rectification (i.e., rectification of quantum states) is the ultimate goal of this research field, and has not yet been realized. We here suggest that the present system could be investigated in the fully quantum regime (considering quantum states for the input light field) as a strong candidate to photonic rectification.


\medskip

{\it Acknowledgements: }
F.F., E.M., D.V., and M.F.S. acknowledge financial support from Conselho Nacional de Desenvolvimento Cient\'ifico e Tecnol\'ogico (CNPq).
A.A. and J-P.P. acknowledge support from Agence National de la Recherche (WIFO project).
D.G. acknowledges the Italian Ministry of University and Research (MIUR) through FIRB - Futuro in Ricerca Project No. RBFR12RPD1, 
and CNPq through the PVE/CSF Project no. 407167/2013-7.
L.S. acknowledges partial support by the Research Council for Natural Sciences and Engineering of the Academy of Finland.
The research leading to these results has received funding from the People Programme (Marie Curie Actions) of the European Union's Seventh Framework Programme (FP7/2007-2013) under REA grant agreement n° [291734].

\medskip

\renewcommand{\figurename}{Fig. S.\hspace{-0.1cm}}
{\it\bf Supplemental material: }\\
{\raggedleft\it Derivation of the reflectance of a TLS:}\\
We take Eq. (1) of Ref. \cite{Daniel} and we adapt it to our case-study. To do this, we set dephasing ($\gamma^*$) and incoherent pump ($\xi$) equal to zero, while efficiency coefficient ($\beta$) equal to unity. We moreover relabel $\delta_L\to\delta\omega$, in order to match the notation used in the Letter. We solve in the steady-state regime, i.e. for $\dd\inn{\sigma_-}/\dd t=\dd\inn{\sigma_z}/\dd t=0$. The solutions are
\begin{equation}
\tag{S.1}
\label{eq:SSsol}
\begin{array}{l}
\Re{\inn{\sigma_-}}= - \frac{\gamma\Omega}{\gamma^2+4\delta\omega^2+2\Omega^2}
\quad,\quad \Im{\inn{\sigma_-}=\frac{2\delta\omega}{\gamma}\Re{\inn{\sigma_-}}}~,\\
\inn{\sigma_z}=-\frac{1}{2}+\frac{\Omega^2}{\gamma^2+4\delta\omega^2+2\Omega^2}~,
\end{array}
\end{equation}
where $\Omega=\gamma\sqrt{2p}$, $\Re$ and $\Im$ are real and imaginary parts, respectively. By plugging this result in Eq. (4) of Ref. \cite{Daniel}, we get $\mathcal{R}=\frac{\gamma^3p}{\gamma^2+4\delta\omega^2+4\gamma^2p}$. Finally, we normalize by the number of photons per second impinging onto the TLS, i.e. $p\gamma$. By doing so, we obtain the reflectance showed in Eq. (1) of the Letter: $R=\frac{\gamma^2}{\gamma^2+4\delta\omega^2+4p\gamma^2}$. A similar calculation is plainly carried out in Ref. \cite{thesis}, where the reflectance is derived and displayed in Eq. (1.51). Using elementary algebra, this latter can be shown to coincide with the reflectance here derived.

\medskip

\begin{figure}[b]
\centering
\includegraphics[scale=0.5]{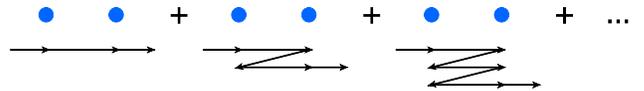}
\caption{(color online). Sum of the amplitudes of events that lead to transmission.}
\label{fig:FigS1}
\end{figure}

{\raggedleft\it Equations for the Fabry-Perot model: }\\
Here we provide explicit derivations of the equations of the Fabry-Perot (FP) model used in the Letter.
Let us consider Fig. 1 of the Letter. Let us furthermore denote by $t_{k>0}^{out}(z)$ the fully transmitted amplitude for the whole FP interferometer at the point $z\ge L$. Such a quantity can be calculated by coherently summing the amplitudes of the events that lead to transmission (see \cite{Yariv}), as sketched in Fig. S.\ref{fig:FigS1}. If we take the phase of the field to be $0$ at the point $z=L$, we get
\begin{equation}
\tag{S.2}
\label{eq:t1}
\begin{array}{lcl}
t_{k>0}^{out}(z)&=&\sqrt{p_{inc}}e^{ik(z-L)}\Big(\sqrt{T_1}\sqrt{T_2}\\[0.4cm]
&&\hspace{-1cm}+\,\sqrt{T_1}\sqrt{R_2}\sqrt{R_1}\sqrt{T_2}e^{2i k L}e^{i(\theta_1+\theta_2)}\\[0.4cm]
&&\hspace{-1cm}+\,\sqrt{T_1}\big(\sqrt{R_2}\big)^2\big(\sqrt{R_1}\big)^2\sqrt{T_2}e^{4i k L}e^{2i(\theta_1+\theta_2)}\,+\,...\,\Big)\\[0.4cm]
&=&\ds \frac{\sqrt{p_{inc}}e^{ik(z-L)}\,\sqrt{T_1T_2}}{1-\sqrt{R_1R_2}e^{2ikL}e^{i(\theta_1+\theta_2)}}~,
\end{array}
\end{equation}
where $T_{1(2)}=1-R_{1(2)}$ is the transmittance of the TLS1 (TLS2) and $k=n\omega/c$. On the other hand, the incident amplitude at the point $z\le 0$, $t_{k>0}^{inc}(z)$, can be written as
\begin{equation}
\tag{S.3}
\label{eq:tinc}
\begin{array}{lcl}
t_{k>0}^{inc}(z)&=&\sqrt{p_{inc}} \; e^{ik(z-L)}~.
\end{array}
\end{equation}
In Eqs. \eqref{eq:t1} and \eqref{eq:tinc}, the subscripts $k>0$ and $k<0$ indicate that light is directed forwardly and backward, respectively.
The fraction of light power that the FP interferometer transmits, i.e. the FP transmittance, can be calculated as 
\begin{equation}
\tag{S.4}
\label{eq:T2}
T=\frac{|t_{k>0}^{out}(L)|^2}{|t_{k>0}^{inc}(0)|^2}=\ds \frac{1}{F_1+F_2\sin^2(2\mu+\theta_+)}~,
\end{equation}
where
\begin{equation}
\tag{S.5}
F_1=\frac{(1-\sqrt{R_1R_2})^2}{(1-R_1)(1-R_2)}\;,\;
F_2=\frac{4\sqrt{R_1R_2}}{(1-R_1)(1-R_2)}\;,
\end{equation}
while $\mu$ and $\theta_+$ are defined in the Letter.

In order to use Eq. \eqref{eq:T2}, we need first to find what the values for $R_1$ and $R_2$ are. This reduces to the question of finding what the values for $p_1$ and $p_2$ are. Similarly to what we have done in Eq. \eqref{eq:t1}, we can calculate the intracavity amplitude for forward ($t^{intr}_{k>0}$) and backward ($t^{intr}_{k<0}$) directions, at the point $z$ ($0\le z\le L$):
\begin{align}
t_{k>0}^{intr}(z)=&\sqrt{p_{inc}}e^{ik(z-L)}\Big(\sqrt{T_1}\,+\,\sqrt{T_1R_1R_2}e^{2i k L}e^{2i\theta_+}\nonumber\\[0.3cm]
&+\,\sqrt{T_1}\big(\sqrt{R_2R_1}\big)^2 e^{4i ( k L + \theta_+)}\,+\,...\,\Big)\nonumber\\[0.3cm]
=&\ds \frac{\sqrt{p_{inc}}\sqrt{T_1}}{1-\sqrt{R_1R_2}e^{2i(kL+\theta_+)}}e^{ik(z-L)}~,\allowdisplaybreaks\nonumber\\[0.4cm]
t_{k<0}^{intr}(z)=&\sqrt{p_{inc}}e^{-ik(z-L)}\Big(0\,+\,\sqrt{T_1R_2}e^{i(kL+\theta_2)}\nonumber\\[0.4cm]
&+\,R_2\sqrt{T_1R_1}e^{i(3kL+2\theta_++\theta_2)}\,+...\,\Big)\nonumber\\[0.4cm]
=&\tag{S.6}\label{eq:tintr}
\ds \frac{\sqrt{p_{inc}}\sqrt{T_1R_2}}{1-\sqrt{R_1R_2}e^{2i(kL+\theta_+)}}e^{i(kL+\theta_2)}e^{-ik(z-L)}~.
\end{align}
Making use of Eqs. \eqref{eq:t1}, \eqref{eq:tinc} and \eqref{eq:tintr}, the light power at the sites of the TLSs, $p_{1,2}$, are obtained by numerically solving the coupled equations 
\begin{equation}
\tag{S.7}
\label{eq:pvalues}
\left\{
\begin{array}{lcl}
p_1&=&\bigg|t_{k>0}^{inc}(0)+t_{k<0}^{intr}(0)\bigg|^2~,\\[0.4cm]
p_2&=&\bigg|t_{k>0}^{intr}(L)\bigg|^2~.
\end{array}
\right.
\end{equation}
The light powers $p_{1,2}$ appear both at the right- and the left-hand sides of Eqs. \eqref{eq:pvalues} in a non-trivial manner. This endows the FP interferometer with an intrinsic non-linear behaviour. These non-linear phenomena are investigated in the Letter.
Equations \eqref{eq:pvalues} depend on all variables that can be externally set in the FP interferometer, viz. $\gamma_{1,2}$, $\delta\omega_{1,2}$, $L$, $p_{inc}$. By using Eqs. \eqref{eq:pvalues}, \eqref{eq:T2} and Eq. \ref{eq:RQDs} of the Letter, the transmittance $T$ can be numerically calculated for any set of those variables.

\medskip

{\raggedleft\it ${\bm\pi}$-shifted intracavity field at the site of TLS1: }\\
Here we give details on the fact that the intracavity field turns out to be $\pi$-shifted with respect to the incident field, at the site of the TLS1, for the settings chosen in Fig. 2 of the Letter. To this regard, from \eqref{eq:tinc},  for $L=0$ the phase of the incident field at $z=0$ is $ik(z-L)=0$. On the other hand, from Eq. \eqref{eq:tintr} the phase of any addend at $z=0$ is $-i\pi$. Thus, the light back reflected from the TLS2 is $\pi$ shifted with respect to the incident light, at the site of the TLS1. This causes zero light intensity at the site of the TLS1, for low light power, as shown in Fig. 3 of the Letter. We may also observe that, when the incident power increases above $\sim 1$, the $\pi$-shifted back reflected light from the TLS2 will not be enough to cancel out the incident light. Therefore the intensity $p_1$ starts growing when $p_{inc}$ approaches $1$.

\begin{figure}[b]
\centering
\includegraphics[scale=0.3,clip=true,trim=0cm 0cm 0.19cm 0cm]{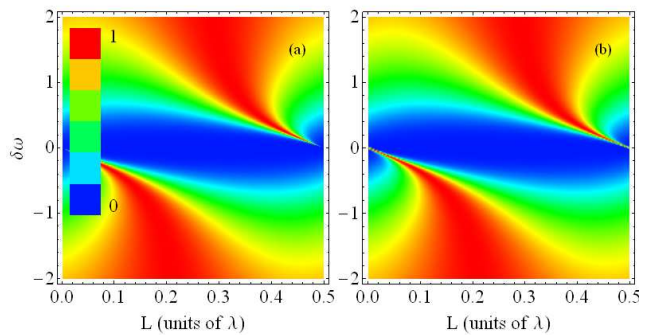}
\caption{(color online). Transmittance for two equal TLSs. a) Quantum mechanical calculation with a single-photon state, as in \cite{Bar2013} and \cite{Shen2005}; b) Calculation from our semi-classical Fabry-Perot model with $p_{inc}=0.001$. The two calculations give consistent results.
}
\label{fig:FigS2}
\end{figure}
%

\medskip

{\raggedleft\it Validation of the semiclassical solution: }\\
We start by analyzing the portion of transmitted light in the case of low incident power and equal TLSs, to directly compare with Refs. \cite{Shen2005, Bar2013, Lal2013}, where this case is considered within quantum mechanical models. We denote $\delta\omega\equiv\delta\omega_1=\delta\omega_2$. In \cite{Shen2005,Bar2013}, the time-independent Schr\"{o}dinger equation for the single photon state is solved. 
For the case of two equal TLSs, \cite{Shen2005,Bar2013} obtain the transmittance showed in Fig. S.\ref{fig:FigS2}(a). On the other hand, results provided by our semi-classical FP model, for the same system, are shown in Fig. S.\ref{fig:FigS2}(b). In this latter, we have chosen very low incident power ($p_{inc}=0.001$), in order to best reproduce the single-photon regime in \cite{Bar2013, Shen2005} by using classical incoming light. We readily see that our semi-classical FP approach in low input regime perfectly reproduces the light transport found in the literature in the single photon regime.


\end{document}